\documentclass{llncs}

\usepackage[svgnames, x11names]{xcolor}

\usepackage[normalem]{ulem} 

\usepackage{listings}

\lstdefinelanguage{XML}
{
basicstyle=\ttfamily\footnotesize,
  morestring=[b]",
  moredelim=[s][\bfseries\color{Maroon}]{<}{\ },
  moredelim=[s][\bfseries\color{Maroon}]{</}{>},
  moredelim=[l][\bfseries\color{Maroon}]{/>},
  moredelim=[l][\bfseries\color{Maroon}]{>},
  morecomment=[s]{<?}{?>},
  morecomment=[s]{<!--}{-->},
  commentstyle=\color{gray},
  stringstyle=\color{blue},
  identifierstyle=\color{red}
}
\usepackage[pdftex]{graphicx}
\graphicspath{{./figures/}}
\DeclareGraphicsExtensions{.pdf}

\usepackage[cmex10]{amsmath}
\usepackage{amssymb}
\usepackage{mathtools}

\usepackage[caption=false]{subfig}

\usepackage{algorithmicx}
\usepackage{algpseudocode}
\usepackage[ruled]{algorithm}
\definecolor{light-gray}{gray}{0.75}
\algrenewcommand{\algorithmiccomment}[1]{\hskip3em{{\footnotesize \textcolor{light-gray}{$\blacktriangleright$}}} #1}

\usepackage{multirow}

\usepackage{xspace}
\usepackage[nocompress]{cite}

\usepackage{enumitem}

\usepackage[english]{babel} 
\usepackage{blindtext}

\newcommand{\uid}{\emph{Aadhaar}\xspace}

\begin{document}
\mainmatter              
\title{Benchmarking Fast-Data Platforms for the \uid Biometric Database}
\titlerunning{Cool Paper}  
%
\author{Yogesh Simmhan, Anshu Shukla and Arun Verma
}
\authorrunning{Simmhan, et al.} 
%
%

\institute{Indian Institute of Science, Bangalore India\\
\email{simmhan@serc.iisc.in, shukla@ssl.serc.iisc.in, arun.verma100@gmail.com}
}

\maketitle              

\begin{abstract}
\uid is the world's largest biometric database with a billion records, being compiled as an identity platform to deliver social services to residents of India. \uid processes streams of biometric data as residents are enrolled and updated. Besides $\sim1$~million enrollments and updates per day, up to $100$~million daily biometric authentications are expected during delivery of various public services. These form critical Big Data applications, with large volumes and high velocity of data. Here, we propose a stream processing workload, based on the \uid enrollment and Authentication applications, as a Big Data benchmark for distributed stream processing systems. We describe the application composition, and characterize their task latencies and selectivity, and data rate and size distributions, based on real observations. We also validate this benchmark on Apache Storm using synthetic streams and simulated application logic. This paper offers a unique glimpse into an \emph{operational} national identity infrastructure, and proposes a benchmark for ``fast data'' platforms to support such eGovernance applications.\footnote{\textit{Pre-print : To appear in WBDB proceedings (LNCS )}}
\end{abstract}
\section{Introduction}
The Unique Identification Authority of India (UIDAI) manages the national identity infrastructure for India, and provides a biometrics-based unique $12$-digit identifier for each resident of India, called \uid (which means \emph{foundation}, in Sanskrit). \uid was conceived as a means to identify the social services and entitlements that each resident is eligible for, and ensures transparent and accountable public service delivery by various government agencies. 

The scope of UIDAI is itself narrow. It maintains a database of unique residents in India, with uniqueness guaranteed by their 10 fingerprints and iris scan; assigns a 12 digit \uid number to the resident; and as an authentication service, validates if a given biometric matches a given \uid number by checking its database. Other authorized government and private agencies can use this UIDAI authentication service to ensure that a specific person requesting service is who they are, and use their \uid number as a primary key for determining service entitlements and delivery. However, to guarantee the privacy of residents, UIDAI does not permit a general lookup of the database based on just the biometrics, to locate an arbitrary person of interest.

\emph{UIDAI holds the world's largest biometric repository.} As of writing, it has enrolled $981$~M of the $1,211$~M residents of India in its database\footnote{India 2011 Census, and live statistics from https://portal.uidai.gov.in}. It continues to voluntarily register pending and newly eligible ones, at an operational cost of about USD~$1$ per person. It also currently performs up to $2$~M authentications \emph{per day} to support a few social services, and this is set to grow to $100$~M per day as the use of \uid becomes pervasive. The \uid database is $5\times$ larger than the next publicly-known biometric repository, the US Homeland Security's OBIM (VISIT) program, which stores $176$~M records on fingerprints, and processes $40\times$ fewer transactions at $88.73$~M authentications \emph{each year} (or $245$~K \emph{per day}), as of latest data available from 2014~\cite{obim:2015}.

Clearly, the \uid repository offers a unique Big Data challenge, both in general and specially from the \emph{public sector}. The present software architecture of UIDAI is based on contemporary open source enterprise solutions that are designed to scale-out on commodity hardware~\cite{arch:uidai:2014}. Specifically, it uses a Staged Event-Driven Architecture (SEDA)~\cite{welsh:sosp:2001} that uses a publish-subscribe mechanism to coordinate the execution flow of logical application stages over batches of incoming requests. Within each stage, Big Data technologies \emph{optimized for volume}, such as HBase and Solr, and even traditional relational databases like MySQL, are used. As part of the constant evolution of the UIDAI architecture, one of the goals is to reduce the latency time for enrollment of new residents or updation of their details, that takes between $3$--$30$ days now, to something that can be done interactively. Another is to ensure the scalability of the authentication transactions as the requests grow to $100$'s of millions per day, and evaluate the applicability of emerging Big Data platforms to achieve the same.

Distributed stream processing systems such as Apache Storm and Apache Spark Streaming have gained traction, off late, in managing the data velocity. Such ``fast data'' systems offer dataflow or declarative composition semantics and process data at high input rates, with low latency, on distributed commodity clusters. In this context, this paper proposes benchmark workloads, motivated by realistic national-scale eGovernance applications, to evaluate the \emph{quality of service} and the host \emph{efficiency} that are provided by \emph{Fast Data} platforms to process high-velocity data streams. 

More generally, this paper highlights the growing importance of eGovernance workloads rather than just enterprise or scientific workloads~\cite{bigdata:nist:2013}. As emerging economies like China, India and Brazil with large populations start to digitize their governance platforms and citizen-service delivery, massive online applications can pose unique challenges to Big Data platforms. For e.g., technology challenges with the \texttt{HealthCare.gov} insurance exchange website in the US to support the Affordable Care Act are well known\footnote{Healthcare.gov: CMS Has Taken Steps to Address Problems, but Needs to Further Implement Systems Development Best Practices, www.gao.gov/products/GAO-15-238}. There are inadequate benchmarks and workloads to help evaluate Big Data platforms for such public sector applications. This work on characterizing applications for the identity infrastructure of the world's largest democracy is a step in addressing this gap. 

The rest of the paper is organized as follows: in \S~\ref{sec:background}, we provide context for the UIDAI data processing workloads; 
in \S~\ref{sec:enroll} and \S~\ref{sec:auth}, we describe the composition of the \uid enrollment and authentication dataflows, respectively. including characteristics of their input stream data sizes and rates; in \S~\ref{sec:experiments}, we evaluate this benchmark for Apache Storm using synthetic streams and tasks based on the real distributions; in \S~\ref{sec:related}, we review related work on Big Data benchmarks; in \S~\ref{sec:discussion}, we discuss how this benchmark can be expanded and generalized, and its relevance on the field; and finally offer our conclusions in \S~\ref{sec:conclusions}.

\section{Background}
\label{sec:background}
UIDAI aims to provide a standard, verifiable, non-repudiable identity for residents of India. This biometric-based identity distinguishes itself from other traditional forms of physical and digital identities in several ways. It is \emph{unique, universal and non-repudiable}, which cannot be said for passports and driving licenses (not universal) or birth certificates and utility bills (not guaranteed to be authentic and unique). \uid guarantees uniqueness of the individual by using their 10 fingerprints and iris scans of both eyes. It is also \emph{electronically verifiable}, since it is just a number and the associated biometric of a person -- it is not a ``physical'' identity card that can be lost, stolen or duplicated. Further, unlike digital identities like OAuth or OpenID, it also proves \emph{physical presence} since the biometric has to be provided by the individual for authentication. Lastly, the fact that it has covered over $80\%$ of eligible residents of India (5 years and older) makes it as close to universal as possible for a voluntary national identity program at this scale.

UIDAI offers two categories of services, one to enroll residents' demographic and biometric data into \uid, and maintain them up to date; and another to authenticate users who provide their \uid number and a biometric. We briefly offer context here, and in the next two sections, we drill down into the workloads themselves. Additional background can be found elsewhere~\cite{arch:uidai:2014}.

\subsection{Enrollment and Update} Residents who have not enrolled into \uid register themselves with an authorized enrollment agency, and provide their \emph{demographic details} (i.e., Name, Date of Birth, Address, EMail, Mobile Number) and their \emph{biometrics} (photo, 10 fingerprints and both iris scans). This data is captured offline (presently), encrypted, digitally signed by the agency, and the \emph{encrypted} enrollment packet for each resident uploaded daily to the UIDAI servers. As part of the enrollment pipeline, the resident's data is validated using basic sanity checks and, importantly, their biometrics is compared against every other resident's biometrics present in the database to ensure duplicate registrations are not performed. This de-duplication provides the authoritative uniqueness to \uid. Once these checks pass, a random and unique 12 digit number is assigned and mailed to the resident. The packets remain encrypted right after data collection, stay encrypted on the network and on disk, and are decrypted in-memory, just in time, during data insertions, queries or comparisons. 

The update process is similar, and allows residents who already have an \uid number to change their transient demographic details, such as address and phone number, or even update their biometrics that could degrade with age. 

\subsection{Authentication and KYC} 
The basic \emph{Authentication service} verifies if a given \uid number matches a biometric or a demographic information provided by a resident, and returns a True/False response. This allows an authorized agency to request verification on whether the individual providing the biometric indeed matches the \uid number they have on record, or if the demographic data that has been provided matches the verified one stored as part of the \uid enrollment. 

The service accepts the \uid number and a combination of the following to be provided -- fingerprint, iris scan, demographic (gender, age, etc.), One Time Personal Identification Number (OTP) -- in an encrypted and signed manner. It then lookups up that \uid number, matches the given biometrics and/or demographics against the details stored for that specific \uid record, and returns a boolean response. The optional use of an OTP ensures that an authentication is done only with the consent of the individual who requests an OTP from UIDAI and provides it to the agency they are authorizing to perform their authentication.

Another \emph{Know Your Client (KYC)} service extends this authentication to also allow the retrieval of demographic data and photo (but not fingerprint or iris scans) by an authorized agency, upon informed consent by the resident. This eases the burden of proof that residents have to provide in order to sign up for public or banking services, and promotes financial inclusion.   



\section{Enrollment Workload}
\label{sec:enroll}
In this section, we describe the proposed fast-data benchmark based on the \uid enrollment application, followed by the authentication application in the next section. As part of the workload, we characterize the dataflow compositions used to process streaming data, the event rate and size distributions that they process, and the required quality of service in terms of end-to-end processing latency.

\subsection{Enrollment Dataflow}

\begin{figure}[t]
  \centering
	\includegraphics[width=\textwidth]{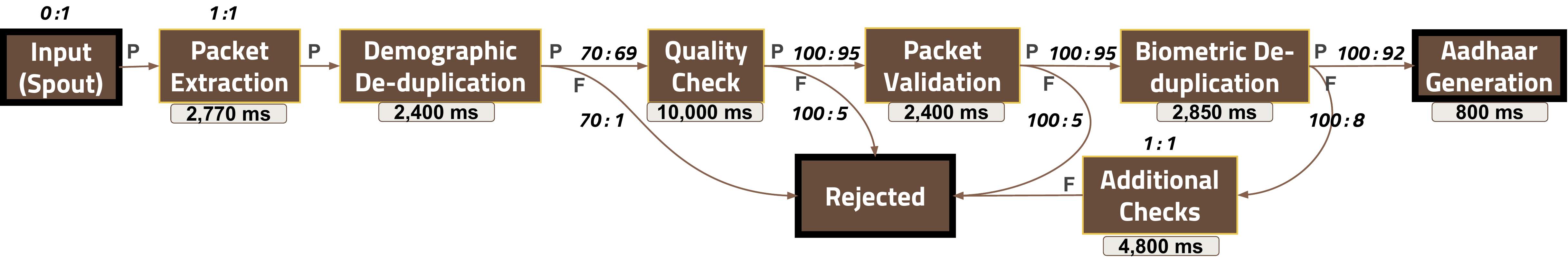}
	\caption{\emph{Enrollment dataflow}. Tasks are labeled with the average latency time in milliseconds. The selectivity is given for each outgoing edge. ``P'' edges are taken by events that pass the check at a task, while ``F'' edges are taken by events that fail a check. 
        }
\label{fig:enroll-dataflow}
\end{figure}
A high level streaming dataflow composition for the Enrollment Workload is shown in Fig.~\ref{fig:enroll-dataflow}. It also shows the \emph{latency} per task to process an input packet, based on observations of the existing application logic, and its \emph{selectivity}, i.e., the ratio between the number of items generated on an output edge for a certain number input items consumed. 

The dataflow starts when encrypted enrollment packets (files) are uploaded to UIDAI by the registering agencies. The input packets' checksums have been verified at a DMZ to detect tampering, duplicate uploads, and malware/virus. The \texttt{Input} task emits such validated packets. Next, the \texttt{Packet Extraction} task decrypts the packets in-memory and inserts the demographics fields and the photo into a MySQL database and a Solr index, and the entire encrypted packet is archived into a distributed, geographically replicated store. 

The \texttt{Demographic De-duplication} task locates existing residents whose demographics, such as name, age and pincode (zipcode), are similar to the incoming packet based on a Solr index search. Candidates that fuzzy-match the demographics have their biometrics preemptively checked against the input packet, and matching input packets, estimated at about $2\%$, are sent to the \texttt{Rejected} task. This avoids a full biometric de-duplication \emph{across all residents} for these inputs.

Then, a \texttt{Quality Check} task does sanity checks on the demographics and the photo to ensure that the photo appears genuine (e.g., based on gender and age), the names and addresses appear valid, detects language transliteration errors during data entry, etc. An estimated $5\%$ of enrollments are rejected by this task. Inputs that pass this QC arrive at the \texttt{Packet Validation} task where its digital signature is verified, and is checked to ensure it was generated by an operator who has been certified, is authorized to operate in that pincode, and matches their assigned supervisor and agency. Typically, $95\%$ of packets pass this check.

Next, the \texttt{Biometric De-duplication} task performs a cross-check of each input packet's iris and fingerprints with every other registered biometric stored in \uid. Three independent Automated Biometric Identification Systems (ABIS) are used here. The input biometric is first inserted into all of them, whereupon they extract a biometric template that they index. Then one of the ABIS is chosen, based on a weighting function, to verify if there is a match for this input from among all indexed biometrics. About $8\%$ of inputs typically find a match and are flagged for potential rejection. Actual rejection of packets depends on \texttt{Additional Checks} of biometric and demographic data, a small number of which require manual checks, before being passed onto the \texttt{Rejected} task.

Packets that pass the prior tasks successfully are sent to the \texttt{Aadhaar\-Generation} task which assigns a unique, random $12$ digit number to this packet, creates a master record for the number, and forwards it for printing and mailing to the resident. Packets that are rejected in any prior step are retained by the \texttt{Rejected} task for auditing in a reject master database.

\subsection{Enrollment Data Stream}
%

\begin{figure}[t]
\centering
  \subfloat[Input Data Size Distribution]{
  \includegraphics[height=1.4in]{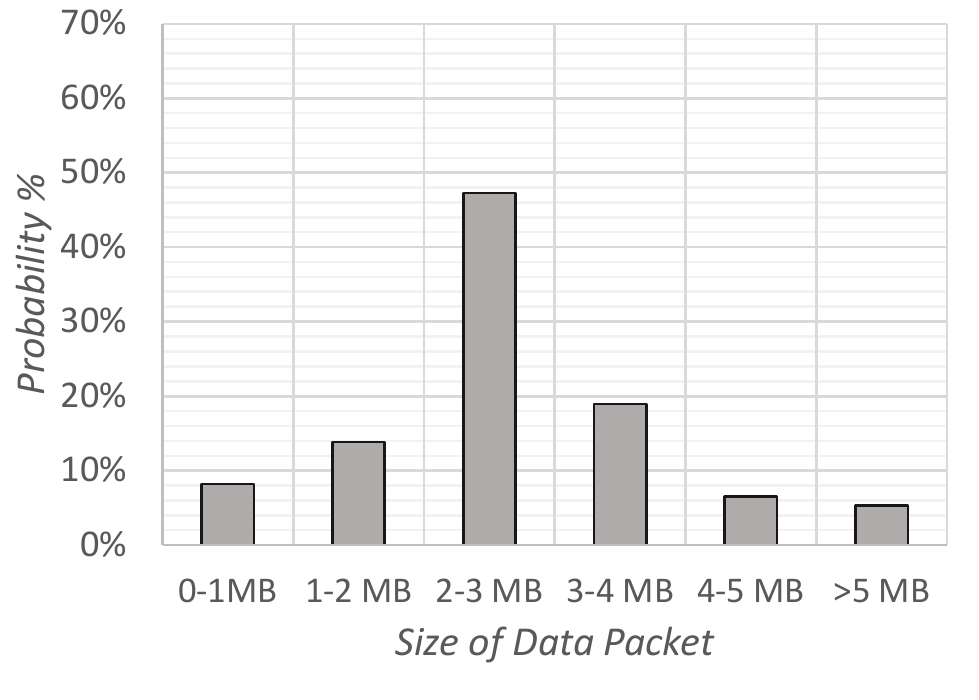}
  \label{fig:enroll-size-distr}
  }
  \subfloat[Hourly Input Data Rate over a $24$~hr day]{
  \includegraphics[height=1.4in]{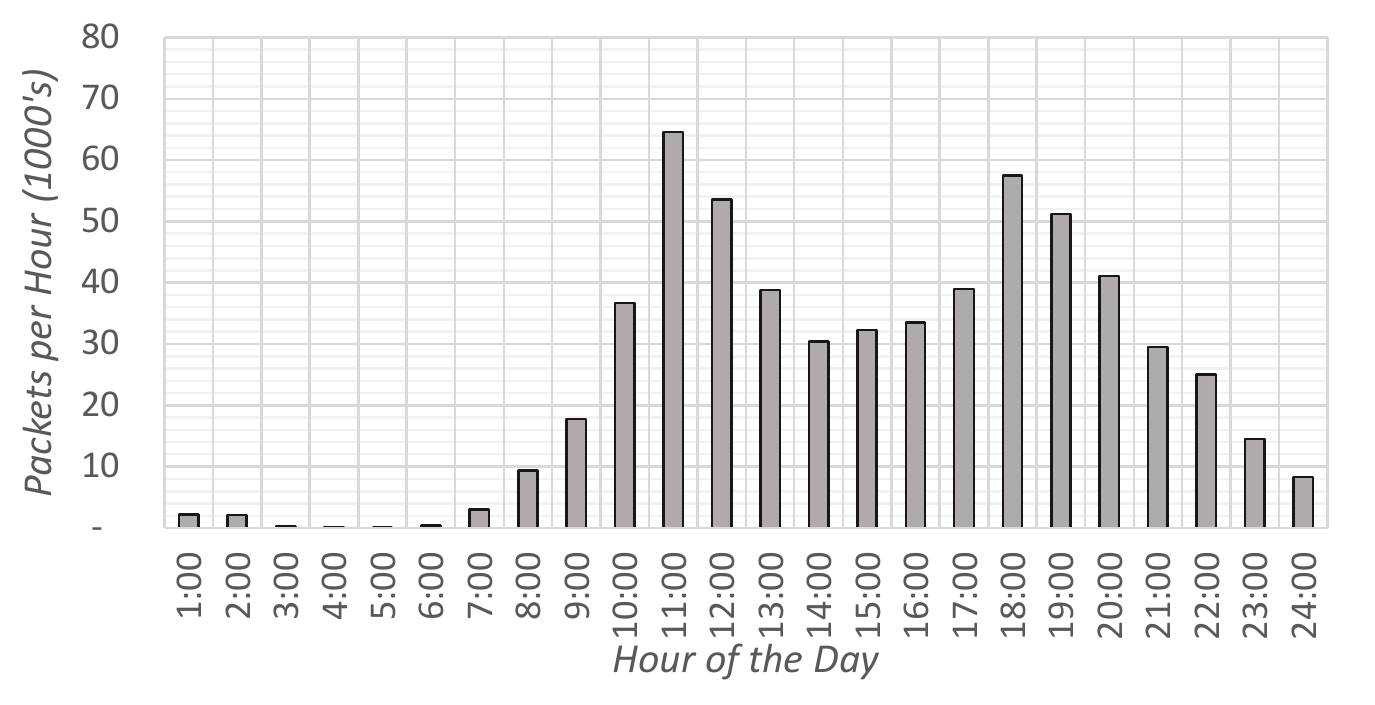}
  \label{fig:enroll-rate-distr}
  }
  \caption{Probability distribution of the input data stream to the \emph{enrollment dataflow}.}
\end{figure}

The transaction rates for the enrollment dataflow change over time, and have to be captured to effectively evaluate the runtime performance of the application. Further, the input data sizes are larger than typical messages to stream processing systems, and hence their variation in sizes needs to be reflected too. Fig.~\ref{fig:enroll-size-distr} shows the probability distribution function (PDF) for the encrypted input enrollment packet size that is uploaded by the enrollment agency and contains the demographics and biometrics. This distribution is from data collected on a single day in September 2015, and is representative of the overall trend. We see that half the packets have a 2--3~MB size, with the overall size ranging from 1--5~MB. We keep the input and output data sizes the same, for the purposes of the workload.

Next, Fig.~\ref{fig:enroll-rate-distr} shows the average hourly input data rates within a recent 24~hour period. This has a bi-modal distribution, with peaks in the late morning and late evening. The rate is low early in the day and by 11AM--12PM, $\sim 65,000$ packets/hour are uploaded in batches by the field agencies after a morning session of registrations. Similarly, enrollment uploads from the afternoon session peaks at 6-7PM. Note that in 2013 when a bulk of the enrollments took place, the enrollment rate was much larger at $1.3$~M packets per day. The output rate from individual bolts and the entire dataflow is determined by the selectivity.

In future, we expect the enrollment agencies to be constantly connected to the Internet, and the enrollment packet uploaded immediately upon capture. So, based on the latency achieved by the streaming enrollment pipeline, an \uid number can be assigned and returned interactively. The input rate distribution is expected to smoothen with such a model.

\subsection{Expected Quality of Service}
Each stream processing workload has a service level agreement (SLA) defined for it based on the end-to-end latency to process a single request. For the enrollment pipeline, the nominal throughput requirement based on a batch-processing model is to complete processing all packets that are received in a 24-hour period, within that 24-hour period. However, given that the advantage of a stream processing approach is to provide lower latency, we define the quality of service expected for processing each packet to be $10$~mins. This would allow residents to eventually enroll for \uid interactively at a service center and be assigned an ID in the same session. 

\section{Authentication Workload}
\label{sec:auth}

\subsection{Authentication Dataflow}
\begin{figure}[t]
  \centering
	\includegraphics[width=\textwidth]{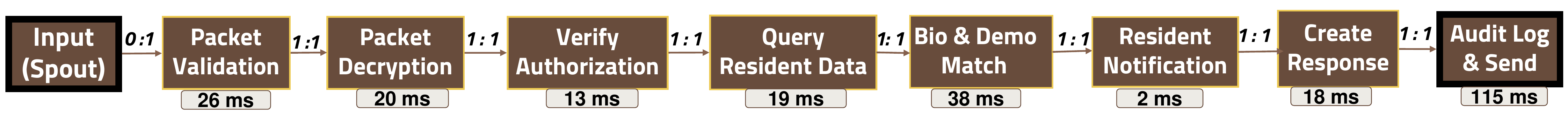}
	\caption{\emph{Authentication dataflow}. Tasks are labeled with the average latency time in milliseconds. Selectivity for all tasks is $1$:$1$.}
\label{fig:auth-dataflow}
\end{figure}

The authentication dataflow composition is shown in Fig.~\ref{fig:auth-dataflow}. This dataflow is pre-dominantly a linear pipeline with a selectivity of $1$:$1$. As before, the latency is shown for each task based on an observational snapshot of the dataflow from Sep 2015.

The \texttt{Packet Validation} stage operates over HTTPS requests that arrive and is responsible for validating the authenticating agency and user, parsing the XML request and verifying the digital signatures of the request. Subsequently, the \texttt{Packet Decryption} task decrypts and extracts the packet's contents and verifies its integrity. The parameters of the request are then parsed by the \texttt{Verify Authorization} task, and a check performed on the type of authentication being done (demographic, biometric) and whether this request is a replay of a previous request. These tasks determine if the request is valid, and is being performed by an authorized entity.

After that, the \uid number present in the request is used by the \texttt{Query Resident Data} task to lookup and retrieve the resident's demographic data from the backend HBase storage. The \texttt{Biometric and Demographic Match} task performs one or more of the following operations based on the type of authentication requested. It checks if an OTP or a PIN number, if present in the request, is valid. It may verify if the retrieved demographic matches the one in the request, if provided. And if a biometric is passed within the request, it checks if this biometric data matches the one stored for that \uid number in the backend ABIS biometric system. These checks determine if the authentication failed or was successful.

As a measure of security, the \texttt{Resident Notification} task asynchronously notifies the owner of the \uid number using their registered email or mobile number that an authentication was performed. Then an XML response is created with the results of the authentication and digitally signed by the \texttt{Create Response} task. Finally, an audit record for the request is created, statistics on the authentication rates and latencies updated for business analytics, and the HTTPS response transmitted to the authenticating agency by the \texttt{Audit Log \& Send} task.

\subsection{Authentication Data Stream}
\begin{figure}[t]
\centering
  \includegraphics[height=1.7in]{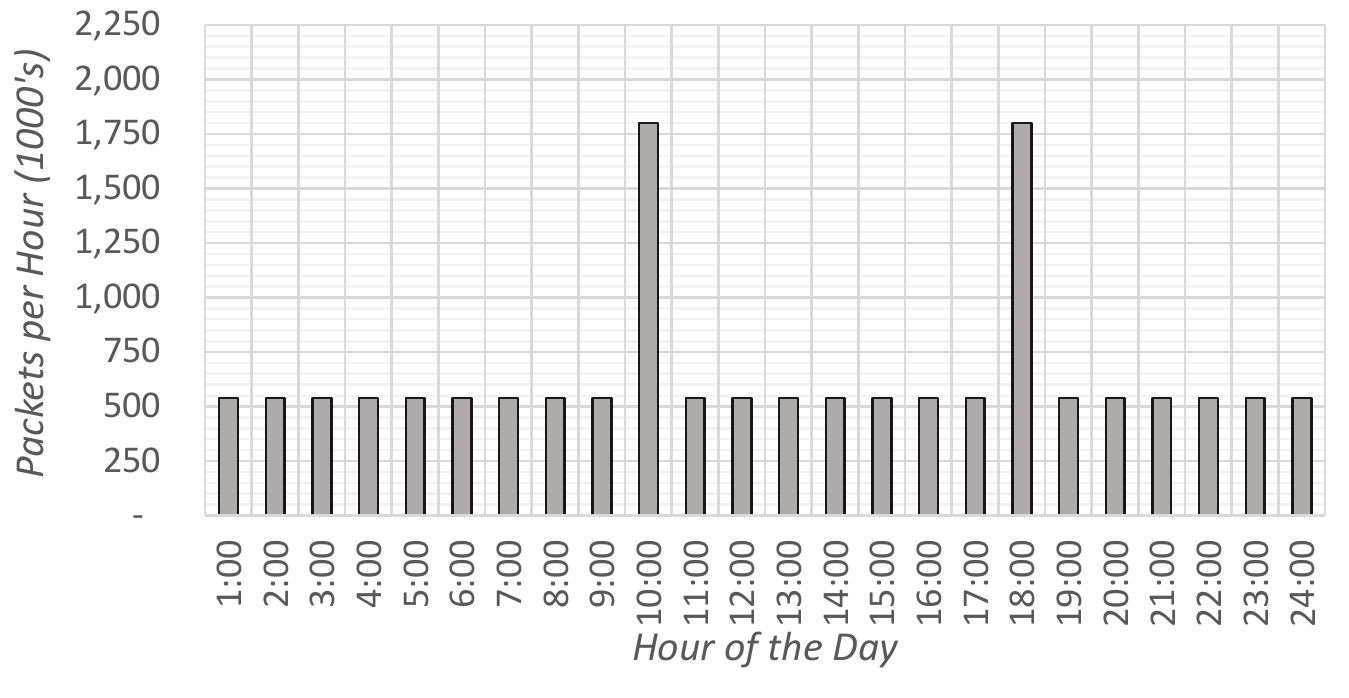}
  \caption{Hourly input request rate over a $24$~hr day passed to the \emph{authentication dataflow} with a base rate of $150$~requests/sec. The two peaks of $500$~requests/sec during the morning and evening periods reflect requests by the federal attendance service.}
  \label{fig:auth-rate-distr}
\end{figure}

%
The input stream to the authentication dataflow is presently more uniform than the enrollment dataflow, even as the number of users performing authentication will increase with time. Input requests are about $4$~KB in size, and we use this constant size per message for our workload. The hourly input data rate is shown in Fig.~\ref{fig:auth-rate-distr}. This is over $20\times$ faster than the enrollment dataflow, and is also expected to grow in future. The current base rate for authentications is $150$~requests/sec, with two sharp peaks of about $500$~requests/sec in the morning and the evening -- \uid is used by federal employees to clock in and out of office each day, and these peaks reflect the requests by this attendance service\footnote{Biometric Attendance Service, http://attendance.gov.in}. In future, the number of authentications are expected to rise to $100$M requests during a working day, or an average of about $2,500$~requests/sec.

\subsection{Quality of Service Expected}
Authentication requests are inherently used by interactive applications. Hence the latency requirements for processing them are tighter. The SLA for end-to-end latency to process a single authentication request by the dataflow is set to $1000$~ms, with most of the requests expected to be completed within $500$~ms. In addition, there are network latencies for transmitting the request and response between the data center and the authenticating agency that add to the round trip time, which we do not consider here.

\section{Experimental Validation}
\label{sec:experiments}
The motivation for the proposed workload is to evaluate the scalability of stream processing systems that can orchestrate the dataflow for the given inputs data streams, and return results within the specified SLAs. Reducing the computational resources required to achieve these quality of service metrics is an additional goal. We implement the proposed workload\footnote{Code and data generator at https://github.com/dream-lab/bigdata-benchmarks} and validate it on the Apache Storm distributed streaming platform.

\subsection{Workload Generation} 

We compose the Enrollment and Authentication pipelines as two \emph{topologies} in Apache Storm. Tasks are defined using a \emph{synthetic bolt} logic that performs in-memory string operations for the given latency duration, and in the process consume CPU resources. The actual workload logic at UIDAI would have performed XML serialization or deserialization, encryption or decryption, remote NoSQL or a relational queries, and so on. Depending on the task, the load on the bolt itself would be limited to string parsing, integer arithmetic (both CPU bound) or waiting for query responses (idle CPU). So, in the absence of access to the actual logic themselves, the string processing performed by the synthetic bolt is a conservative approximation of the expected computational effort taken by each task in the dataflow.

The synthetic bolt is configured for each task to perform its computation for the given \emph{latency} duration for that task. The bolt is also configured with the \emph{selectivity} on each of the output edges for that task, and based on this, a certain fraction of input packets are routed to the downstream bolt(s) on the relevant edges. If the selectivity matches, the bolt logic passes the input packet to the relevant downstream bolt(s) without change.

We have developed an \emph{event stream generator} tool that uses the given input rate and size distributions of the enrollment and authentication packets to pre-generate events with the appropriate relative timestamps and payload sizes. Synthetic events of these sizes are pre-fetched into memory by a \emph{spout} logic within the Storm topology, and replayed at the appropriate relative time intervals to match the required input rate distribution. In particular, the spout uses a multi-threaded and distributed mechanism to ensure that the data rate can be maintained at even 10,000's of events/sec, if necessary. 
As we show, the results confirm that the observed input rates and sizes for the experiments match the reference input distribution that is given. The spout can also be configured with different time scaling factors to speed-up or slow-down the rates while maintaining a proportional inter-arrival time between events. 


\subsection{Deployment} 

We run the topology on a 24-node commodity cluster, with each node having 8 AMD Opteron 2.6~GHz Cores and 32~GB RAM, connected by GigaBit Ethernet, and running Apache Storm~v0.9.4 on OpenJDK~v1.7 and CentOS~7. Storm \emph{supervisors} that execute the tasks of the topology run on 23 nodes and 1 node is dedicated to management services such as Master, Zookeeper and Nimbus. There are eight resource \emph{slots} in each supervisor (host), one per CPU core, on which the Storm \emph{workers} can execute one or more task threads. 

Storm topologies can be configured to use only a subset of the supervisor slots in the cluster and also the degrees of parallelism (number of threads) assigned to each task in the topology. By default, each task runs on a single thread on a single slot. Based on the SLA required for each pipeline and data rate distribution, we determine the \emph{minimum number of slots} that should be allocated to the topology and the \emph{degree of parallelism per task} required to meet the SLA for the benchmarks. 
These configurations are based on the latency time per task in the dataflow such that there are adequate threads to process data arriving at the \emph{peak rate} for that workload, and ensure that there are adequate CPU cores to sustain the compute requirements of these threads, yet without punitive context-switching overheads.

For the enrollment topology, the total degree of parallelism required to meet these latencies and peak rate comes out to $475$ threads, shared uniformly across all the tasks, and running on $9$ nodes ($72$ cores) of the cluster. For the authentication topology, we arrive at a total degree of parallelism of $514$ threads, distributed proportionally across all tasks based on their contribution to the overall latency of the topology, and using $19$ nodes ($152$ cores). The spouts that generate the input streams in parallel for the enrollment and authentication dataflows are included within this count and they take up $1$ thread and $10$ threads, respectively, while the terminal sink tasks in the topologies take up $1$ thread each.

\subsection{Results} 
\subsubsection{Enrollment Workload.}
We perform a $24$-hour benchmark run where data streams that follow the given rate and size distributions are generated and passed as input to the Enrollment topology in Storm for a whole day. Overall, about $591,270$ input packets were generated during this period. 

\begin{figure}[t]
\centering
  \subfloat[Expected and actual input data sizes. Observed rates are grouped into 4-hour periods. We see that all periods have the same distribution.]{
  \includegraphics[height=1.35in]{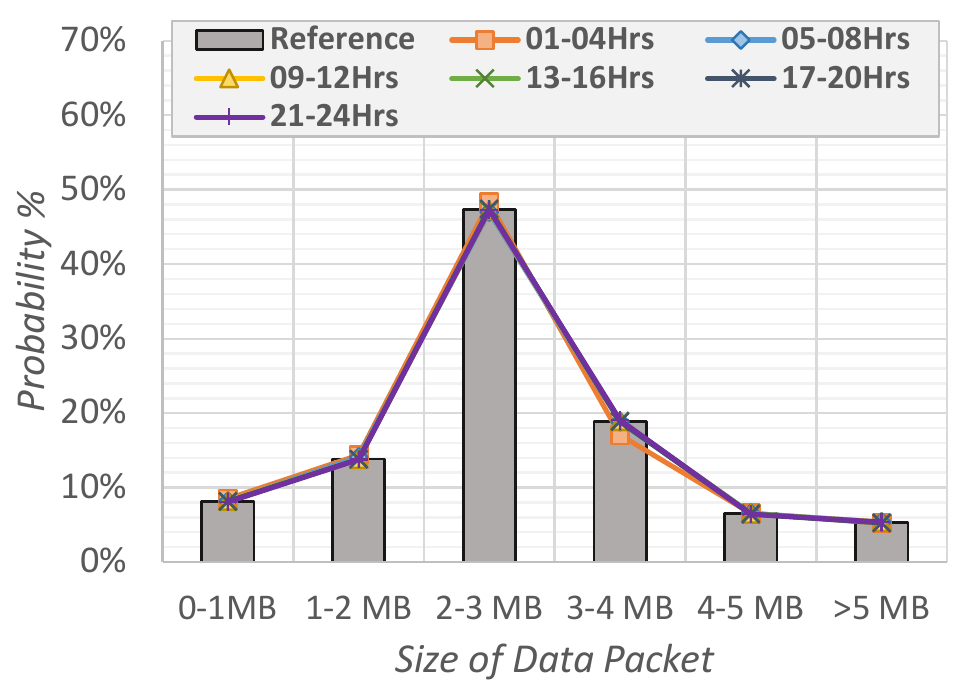}
  \label{fig:enroll-size-actual}
  }~~~
  \subfloat[Expected hourly input rate (bar), and the actual hourly input and output rates (lines) over a $24$~hr period.]{
  \includegraphics[height=1.37in]{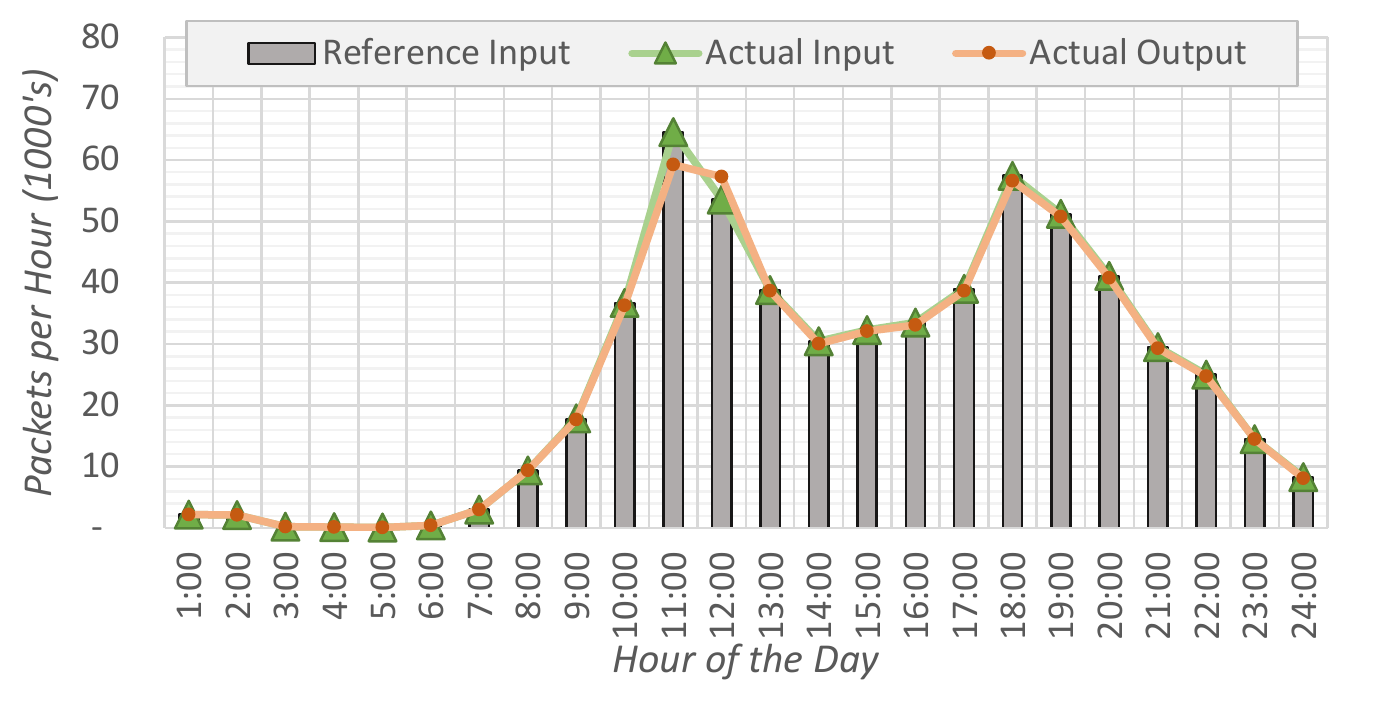}
  \label{fig:enroll-rate-actual}
  }
  \caption{Probability distribution of the expected and observed input data stream to the \emph{enrollment workload} experiment.}
  \label{fig:actual:enroll}
\end{figure}

Figs.~\ref{fig:enroll-size-actual} and \ref{fig:enroll-rate-actual} show characteristics of the expected and actual input stream. We are able to validate that the event generator can generate input events with the same size distributions as the reference size distribution (Fig.~\ref{fig:enroll-size-actual}, gray bars), not just cumulatively for the entire $24$-hour period but also for individual $4$-hour periods (Fig.~\ref{fig:enroll-size-actual}, lines). Likewise, we see from Fig.~\ref{fig:enroll-rate-actual} that the hourly input rates generated by the spouts (green line/triangle) match the reference input rate distribution (gray bars). The figure also shows that the output event rate (orange line/circle) from the topology closely matches the input rates (green line/triangle). The output rate falls behind by $\sim5,000$~packets/hour at $11$~AM, where the input rate peaks, but it is able to compensate for this and catch up within the next two-hour period. 

\begin{figure}[t]
\centering
\subfloat[Violin plot of latency distribution. The minimum theoretical latency for a successful enrollment is $21.22$~secs, shown as a dashed green horizontal line.]{
  \includegraphics[height=1.45in]{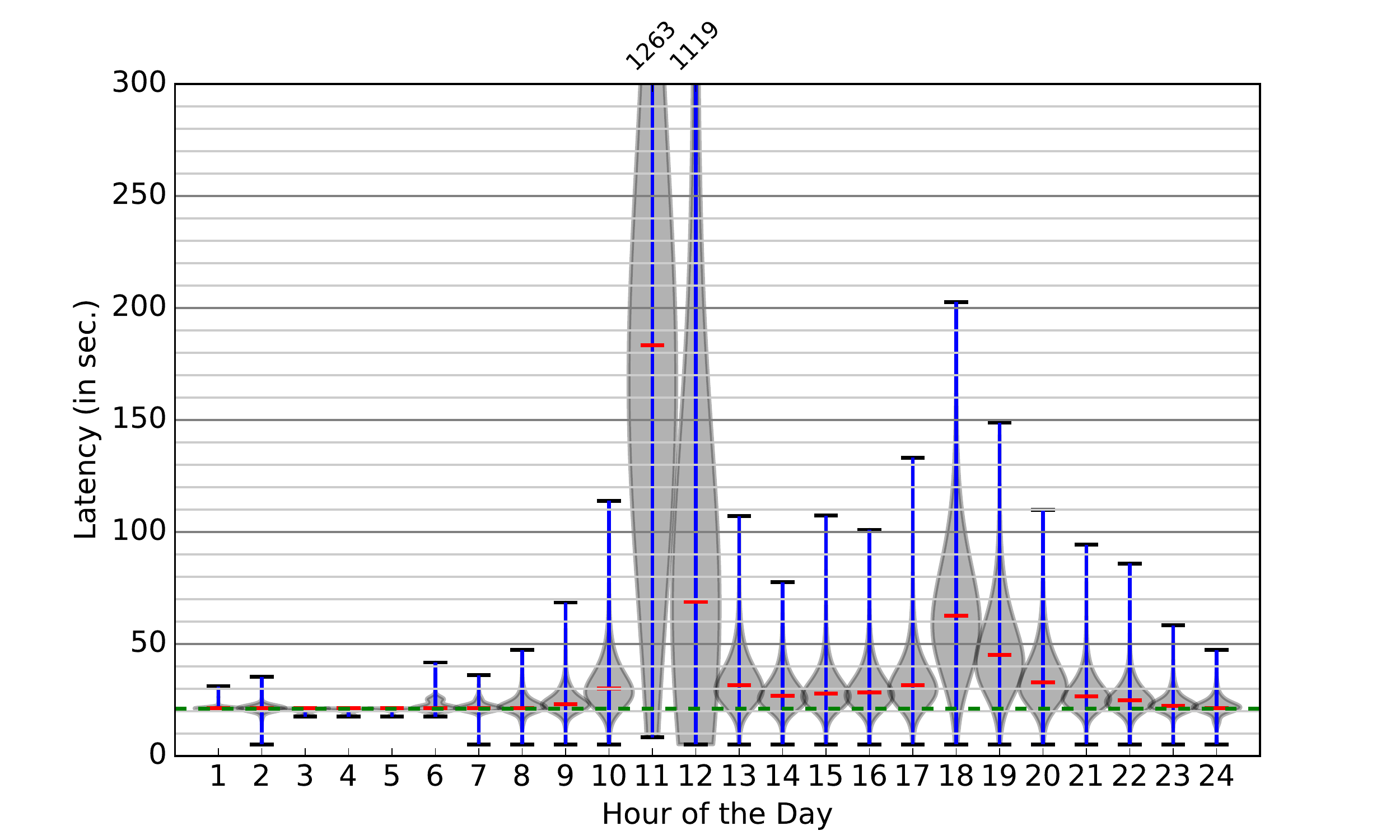}
  \label{fig:enroll-latency}
}~~~
  \subfloat[Violin plot of CPU utilization\% sampled every second for $9$ nodes, for a $1/10^{th}$ time duration experiment that is scaled back to $24$~hours for plotting.]{
  \includegraphics[height=1.4in]{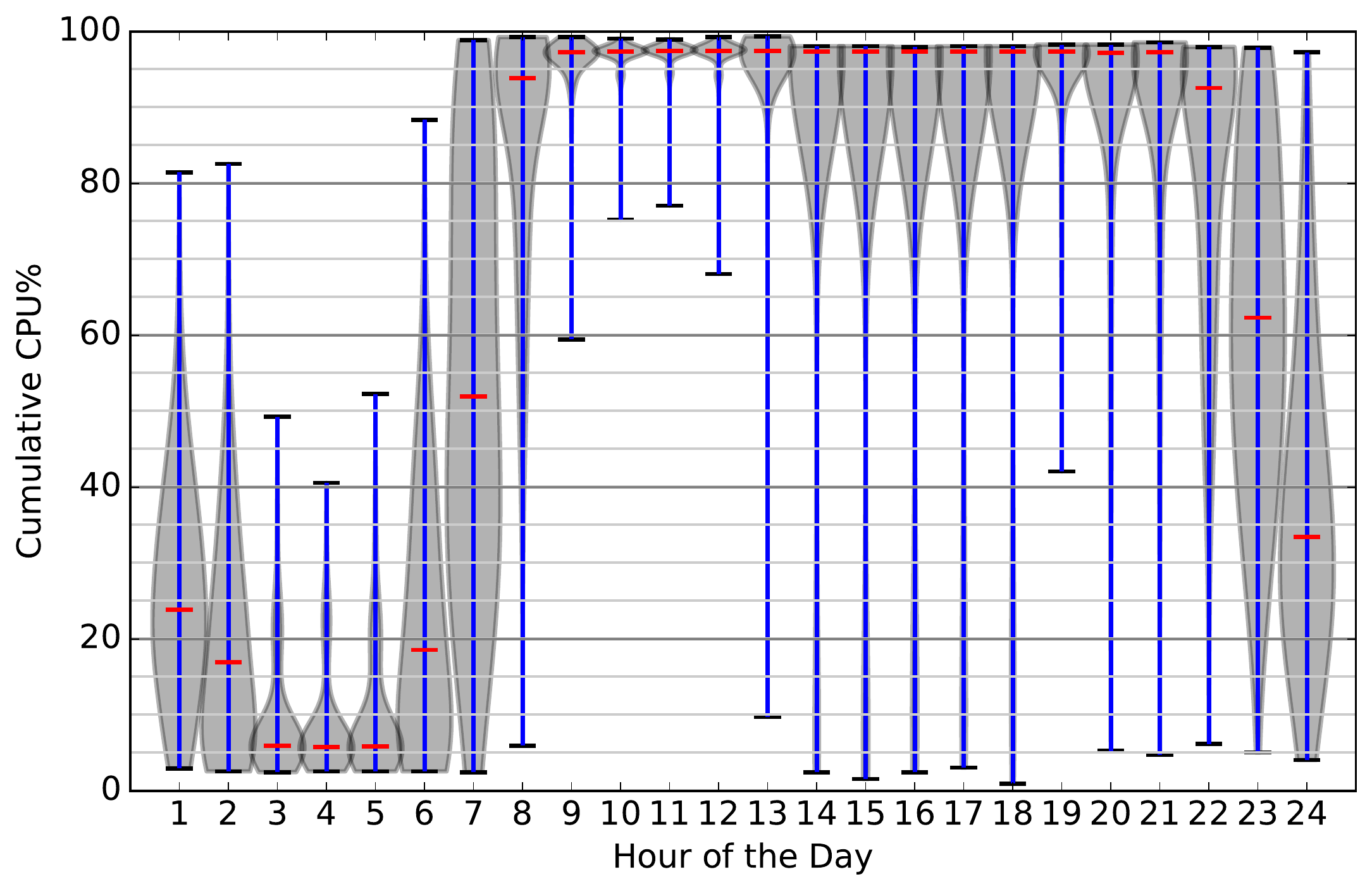}
  \label{fig:enroll-cpu}
  }

  \caption{Violin plot on hourly distribution of \emph{latency} per packet (secs) and \emph{CPU utilization} for the \emph{enrollment workload} experiment.} 
  \label{fig:box:enroll}
\end{figure}

Drilling in further into the \emph{event latencies}, Fig.\ref{fig:enroll-latency} shows a violin plot\footnote{The \emph{violin plot} is a generalization of a box and whiskers plot. The minimum, median and maximum values are marked with a dash on the vertical line. The width of the horizontal shaded region around each vertical bar represents the relative frequency of packets having that latency value. } during each hour for the end-to-end latency for every event packet in that hour. 
Here, we note that the average hourly \emph{median} latency per enrollment packet is $35.74$~secs, relative to the theoretical lower bound makespan of $22.22$~secs per event that excludes the network time and other overheads. Except for the $11$~AM and $12$~PM hours, when we have a peak input rate and the system is catching up, the \emph{maximum} observed latency time for packets during all other hours is less than $202$~secs per packet -- well below the $600$~secs SLA. 

For the $11$~AM and $12$~PM hours, the median latency is $183$~secs and $69$~secs, respectively, with a peak at $1,263$~sec. The evening input rate spike at $6$~PM does not have such a strong impact on the latencies, and they go up only marginally to a median value of $62$~secs and a peak of $202$~secs. In all, $17,700$ events out of a total of $591,270$ events (i.e., $2.99\%$) had an SLA violation, where the end-to-end latency per event was greater than $600$~secs. In fact, \emph{for $97\%$ of the inputs, we are able to complete the enrollment pipeline within just $5$~mins.}

In order to evaluate the \emph{resource efficiency} of Apache Storm in processing the workload, we sample the CPU utilization every second on each of the 9 active nodes for this topology. To keep the logs manageable, this is done for a shorter experiment that ran for $1/10^{th}$ the whole-day duration (i.e., $2$~hours and $24$~minutes), while keeping the same size and rate distributions scaled down in time. Fig.~\ref{fig:enroll-cpu} shows an hourly violin plot of the CPU utilization\% sampled every second per node, with each bar having $(9 \times 60 \times 60 \times \frac{1}{10})$ samples. 
As the pipeline warms up in the initial hour, we see that the CPU utilization is at about $20\%$, and beyond that, we see a correlation of the utilization with the input rate. Once we hit the morning peak corresponding to the $11$~AM period, the CPU utilization is close to $100\%$ for all nodes in the cluster during the entire hour, exhibiting high efficiency. The median utilization remains close to $100\%$ for $13$ straight periods, until the effects of the morning and evening peaks wear off by the $10$~PM period. The average CPU utilization across all nodes for the entire period was high at $70\%$, despite the input rate variation.


\subsubsection{Authentication Workload.}
\begin{figure}[t]
\centering
  \includegraphics[height=1.7in]{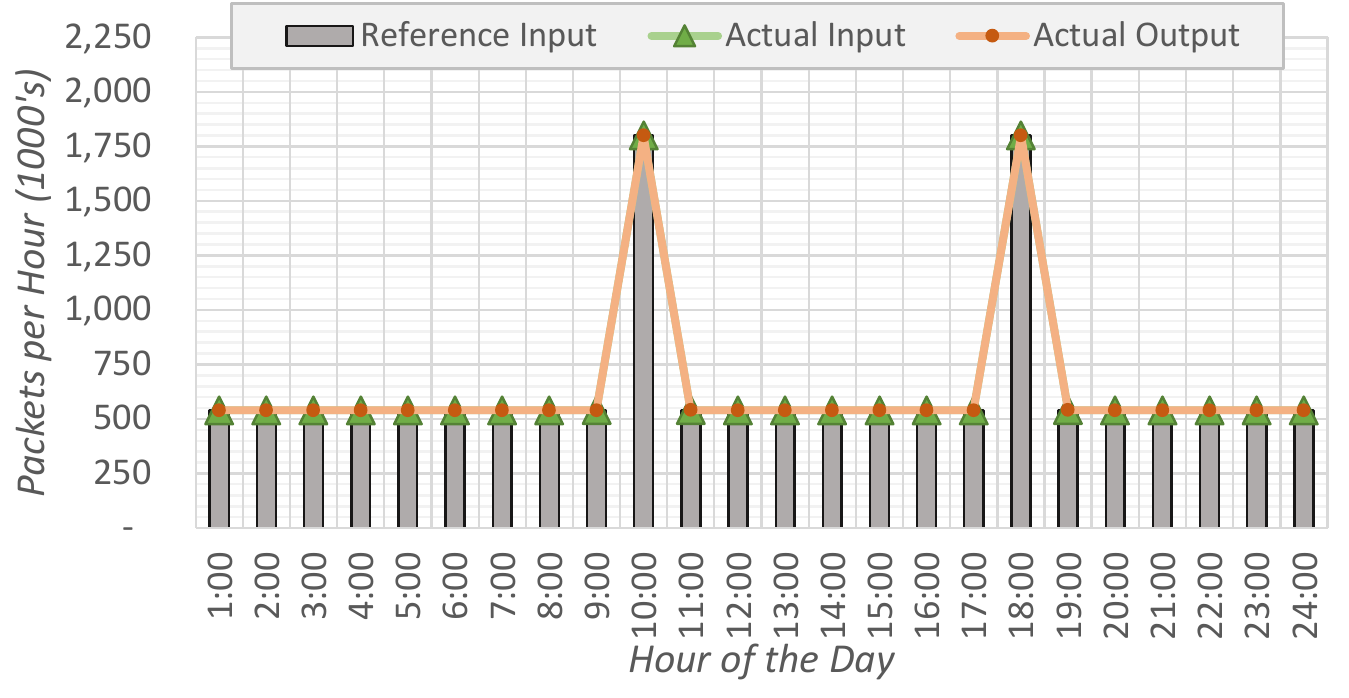}
  \label{fig:auth-rate-actual}
  \caption{Hourly reference and actual input data rates passed to the \emph{authentication dataflow} during a $24$~hr experiment run, and the actual hourly output rates from the dataflow.}
\end{figure}

The event generator for the authentication workload was also run for a $24$-hour period and about $15,480,000$ events are passed to the authentication topology during that time. Fig.~\ref{fig:auth-rate-actual} shows that the reference and actual hourly input rates for the topology. The observed input rates are able to match precisely with the much higher, albeit smaller in size, expected input rates for the authentication workload compared to the enrollment workload. Since one output event is generated by each input event passed to the topology, the figure also shows that the output rate is able to sustain and match the input rate at the hourly granularity, even during the AM and PM spikes.

\begin{figure}[t]
\centering

  \subfloat[Violin plot of latency distribution. The minimum latency for a successful authentication is $250$~ms, shown as a dashed horizontal line.]{
  \includegraphics[height=1.45in]{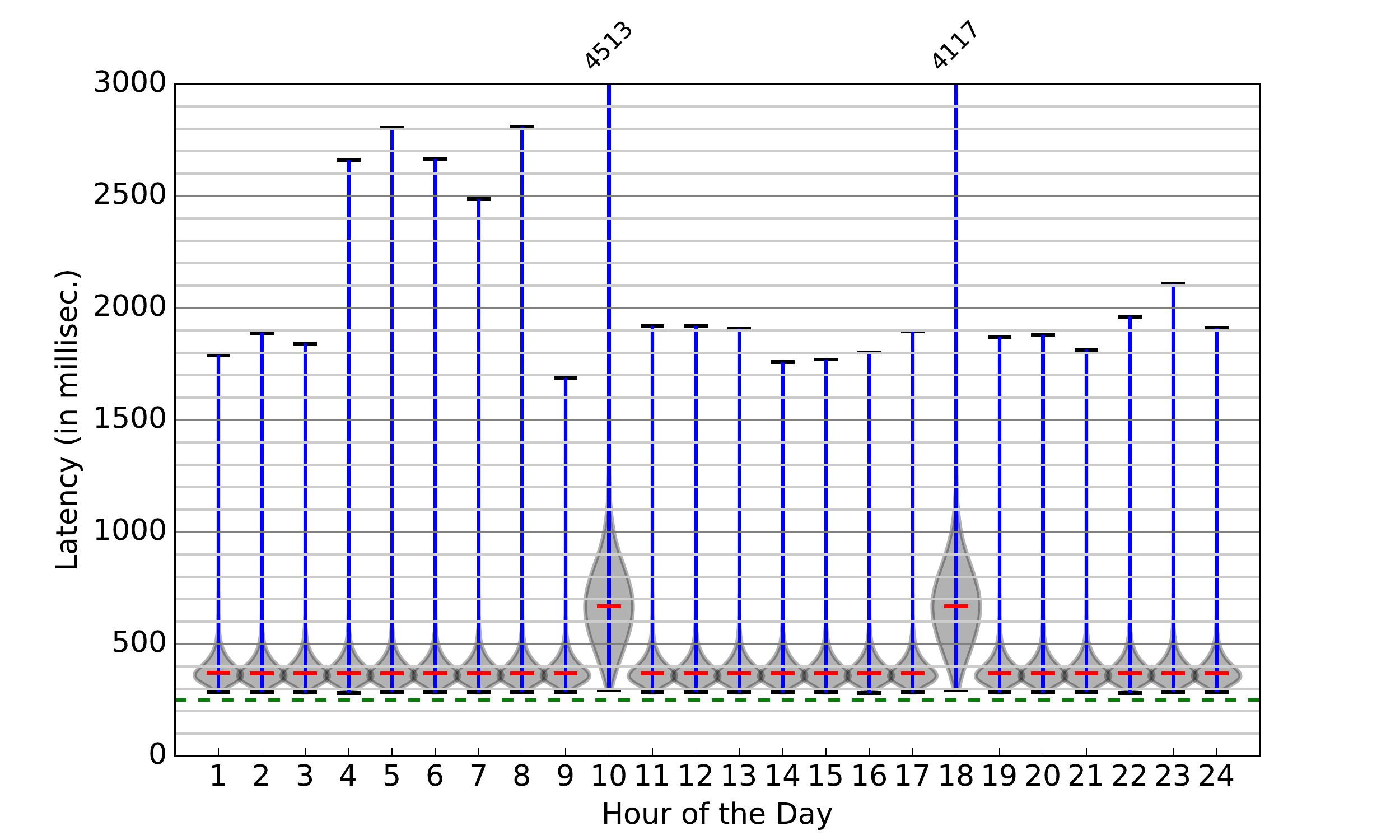}
  \label{fig:auth-latency}
}~~~
  \subfloat[Violin plot of CPU utilization sampled every second for $19$ nodes, for a $1/10^{th}$ time duration experiment that is scaled back to $24$~hours for plotting.]{
  \includegraphics[height=1.4in]{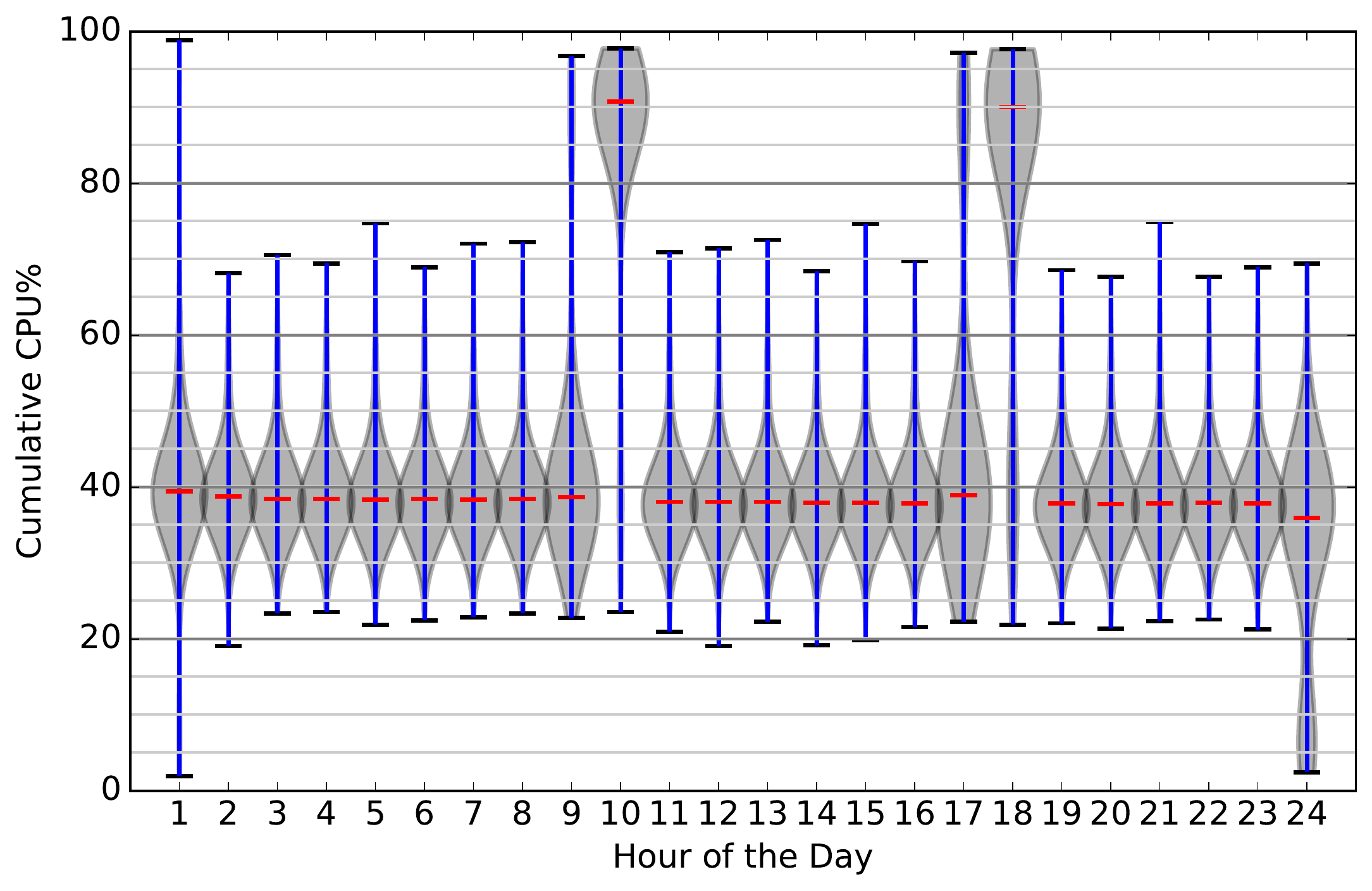}
  \label{fig:auth-cpu}
  }


  \caption{Violin plot of hourly distribution of \emph{latency} per request (ms) and \emph{CPU utilization} for the \emph{authentication workload} experiment.} 
  \label{fig:box:auth}
\end{figure}

We examine the end-to-end event \emph{latency} distributions for every hour using the violin plot shown in Fig.~\ref{fig:auth-latency}. We see that except for the $10$~AM and $6$~PM hours that correspond to the input rate spikes, the bulk of the latencies for all other hours are tightly bounded, and fall predominantly in the range of $350$--$500$~ms. This is in relation to the theoretical lower bound latency of $250$~ms that does not take network costs into account. During the two spikes, the latency both increases and its distribution widens, with the most of the events falling in the latency range of $550$--$1,000$~ms, but within the SLA of $1$~sec. While there are outliers in all hours that sometimes reach $4,500$~ms, overall, we see that \emph{the SLA is met for $99.98\%$ of the input events}, and violated for only $2,430$ events out of over $15$~M.

The violin plot of the \emph{CPU utilization} across $19$ nodes is shown in Fig.~\ref{fig:auth-cpu}, for a shorter experiment that ran for $1/10^{th}$ the period. Unlike the enrollment workload, we see that the median utilization stays low at about $40\%$ for most of the hours except for the $10$~AM and $6$~PM input rate spikes when the median goes up to $93\%$. These utilization values however drop back in the next hour, indicating that the computation load stabilizes quickly after the input rate drops back to the base load of $150$~requests/sec. The average utilization across all hours and nodes is $43\%$. 

Relative to the enrollment workload, a higher fraction of input events fall well within their SLA limits for the authentication workload but we also observe a lower CPU utilization. This indicates that we can potentially trade-off resource efficiency and latency in the case of the authentication dataflow. The resource allocation could potentially be reduced for better efficiency, though the latency distribution for the two peaking hours may come closer to or go beyond the SLA limits. Similarly, the enrollment workload could improve its SLA beyond $97$\% by increasing the resource allocation, but consequently have a resource utilization lower than $70$\%. This also motivates the need for elastic resource allocation over time.




%
\section{Related Work}
\label{sec:related}
Benchmarking distributed stream processing systems (DSPS) using real world workloads can help identify which of their features are impacted by unique input data characteristics, composition capabilities, and latency requirements. Several benchmarks have been proposed in this context.

The \emph{Linear Road Benchmark}~\cite{lrb} simulates a highway toll system for motor vehicles  with variable tolling, and was meant to compare Data Stream Management Systems (DSMS) with Database Management Systems (DBMS). The input to the benchmark is from a traffic model that has variable number of vehicles, but each emitting tuples at a uniform rate and with the same type. So while the input rate is variable, each message is of a fixed, small size. The metrics for evaluation are the response time and accuracy of the query, and the maximum sustained rate within a specified response time, but it does not consider resource utilization. While not developed for distributed stream processing systems, which deal with opaque messages and user logic, it can be adopted to validate DSPS. Our proposed benchmark is particularly tuned for DSPS and simulates the behavior of a real-world eGovernance workload, with variable message sizes and resource efficiency as an additional metric.

\emph{StreamBench}~\cite{lu:ucc:2014} proposes 7 micro-benchmarks on 4 different synthetic workload suites generated from real time web logs and network traffic. Different workload suites are classified keeping performance metrics in mind. Performance workload and Multi-recipient performance workload suites measure the throughput and latency by pushing up the input rate, using single and multi receiver respectively; Fault tolerance workload suite measures the throughput penalty factor by deliberately causing nodes to fail; and Durability workload suite measures the fraction of time for which the framework is available when running the experiments for long durations. While we do not consider durability or fault-tolerance directly, our benchmarks are run for $24$-hours, and this can be extended to longer periods to test durability. In addition to message rate variability, we also consider larger message sizes and variations in message sizes that can impact the resource utilization, which we additionally consider as a metric for comparison between DSPS platforms. 

The \emph{IBM Streams benchmark}~\cite{nabi:ibmstreamsreport:2014} uses a email spam detection application over the Enron email dataset, and does a relative comparison of IBM's Infosphere Streams and Apache Storm. While they are able to reproduce the data sizes (KBs) and rates from real emails from over a decade ago, their dataflow itself does not capture any application logic or even sleep times. So this benchmark is more a measure of the network overheads when running the dataflow, rather than computational resources used by, or the consequent latency of, the dataflow. Hence it has limited value. 

While such stream processing benchmarks are useful, they do not consider message payloads that are large (MBs) in size, provide strict latency SLA requirement based on real applications, or capture characteristics of eGovernance services. Our \uid benchmarks complements these efforts.

\emph{Chronos}~\cite{gu:chronos:2015} is a generic data generation framework for stream benchmarks. Its focus is not to actually benchmark or stream the input data, but to generate data that can then be streamed to perform benchmarking. Chronos can create large scale synthetic time-series input data that mimics the distributions and correlations that it identifies from a given sample data. It uses Latent Dirichlet Allocation (LDA) method for extracting  temporal dependencies from the sample data and preserves correlation among columns. This avoids users having to mine the data for complex patterns and distributions themselves for generating benchmark input specifications. This complements our work, where the distributions are explicitly given based on historic observations, and the focus is on benchmarking streams with different size and rate characteristics over the given dataflow definitions.

Other comprehensive benchmarks have developed for Big Data processing, beyond just fast-data processing which is our emphasis. \emph{Hibench}~\cite{huang:hibench:2010} includes 10 micro workloads covering SQL Queries, Machine learning, Graph Computation and micro-benchmarks like word count and sort. \emph{SparkBench}~\cite{li:scb:2015} comprises of different application types like Graph computation, SQL queries along with streaming application for Spark Streaming. This is a mixed-workload for different dimensions of Big Data. It uses resource consumption and data processing rate as the metrics for evaluation, but processing latency is not evaluated. 
%

\emph{Big Bench}~\cite{ghazal:acm:2013} is modeled on the TPC-DS benchmark~\cite{poess:sigmod:2002} and uses a retail industry scenario. It offers both semi-structured and unstructured data for data variety; analytics queries over click logs and user reviews, contributing to data volume; and an extract-transform-load (ETL) pipeline for data velocity. This is evolving as a community Big Data benchmark and has been validated on Teradata Aster and Hive platforms, among others. Variations to TPC-DS have also been proposed~\cite{zhao:wbdb:2013}. Similar to these Enterprise Big Data workload suites, one can envision an eGovernance Big Data workload suite, with this paper on the velocity dimension being a starting point.




%
\section{Discussion}
\label{sec:discussion}
\subsection{Extending the Workloads}
The fast-data workloads that we have proposed are based on initial observations on the structure and latencies of the two \uid dataflows, and the distributions of their input packet rates and sizes. There are several opportunities to expand upon these, both to increase the value derived from the workloads and to evaluate other Big Data platforms that support these applications.

The tasks in the two dataflows themselves are diverse in terms of the resources that they consume. Our use of a string processing synthetic task is a conservative proxy for the actual tasks. We can further \emph{categorize the tasks in the dataflow based on their resource intensity} into CPU intensive, I/O intensive, memory intensive or idling tasks -- the latter of which runs a remote query but does not consume much resources on the task's local host. Based on this, we can have different categories of simulated tasks be used in the workload which will help to model their resource consumption better. This will offer more genuine estimates of the degrees of parallelism required for each task, service level agreements that can be met, and the resource efficiency possible.

Occasionally, new and better biometric identification algorithms may be considered to improve the robustness of authentication. Then, all prior biometric data collected during enrollment may be \emph{reprocessed} to evaluate the new algorithms. This reprocessing can use a variation of the enrollment pipeline to operate in a bulk mode, and compare the quality of the old algorithm with the new. There are also one-off scenarios when \emph{bulk authentications} are performed for demographic verification by specific agencies, often during off-peak hours at night. Such high-throughput variations to the proposed low-latency workloads are worth considering.



The data rate distributions themselves are bound to change over time, and with \uid's use of multiple regional data centers, the load on the backend may have \emph{region specific trends} that can be captured. It may also be possible to identify rate distributions specific to different authenticating sources, and generate synthetic \emph{cumulative distributions} that blend them in different phases for representative, multi-modal, longer-term simulation runs. These can also help identify extreme spikes that may be possible when different distributions co-incidentally peak at the same time, and help understand the behavior of the system during exigencies.

The trade-off that we observe between resource efficiency and SLA violations also provides the opportunity for developing and evaluating \emph{resource allocation strategies} for fast-data platforms. The rate variation combined with the different probabilities for paths taken through the dataflow means that the load on each task is not uniform across time. Thus, these workloads can be used to benchmark the agility of fast-data platforms to intelligently acquire and release elastic Cloud resources to achieve the SLA while also reducing the (actual or notional) monetary cost for using virtualized resources.

While the stream processing dataflows coordinate the execution of the business logic, there are other high \emph{data volume} workloads that are performed on the backend platforms that actually host the \uid data and respond to queries from the streaming pipelines. These span NoSQL and relational databases that host demographic data, biometric databases that index and query over fingerprint and iris data, distributed file-systems that archive raw enrollment data and results, and audit logs from billions of authentications that are useful for mining. Each of these are a Big Data platform case study in itself, and deserve attention as part of a eGovernance benchmark that cuts across Big Data dimensions.










\subsection{Practical Considerations}
The proposed \uid benchmarks are validated using the Apache Storm distributed stream processing system. Other platforms such as Apache Spark Streaming and InfoSphere Streams could similarly be validated. Our results from evaluating Storm show that such fast-data platforms can achieve the SLAs that are required for the enrollment and authentication dataflows, and have the potential to significantly improve the quality of service for the end user.

However, such a validation of the orchestration platform is just one piece of a complex architecture, and cannot be construed to making an immediate operational impact within \uid. The current SEDA model, which a distributed stream processing system could conceivably replace, is one of many Big Data platforms that work together to sustain the operations within UIDAI. Switching from a batch to a streaming architecture can have far-reaching impact, both positive and negative (e.g., on robustness, throughput), that needs to be understood, and the consequent architectural changes to other parts of the software stack validated. Making any design change in a complex, operational system supporting a billion residents each day is not trivial. 

There are also logistical considerations since the overall application workflow relies on agents on the field and technology availability outside the data center. Some of the tasks in the dataflow, such as \texttt{Quality Check} in the enrollment dataflow, has human agents in the loop who complement automated quality checks. These have to be modeled better to guarantee the notional SLAs we observe, or there should be sufficient confidence in complete automation. Mobile field offices that collect enrollment data do not have Internet connectivity round-the-clock, since they may be at remote villages, and currently upload the data in batches each day. Moving to an interactive enrollment process necessitates constant Internet connectivity, which may be possible in the near future but requires additional resources and planning.

So, in summary, our proposed benchmarks are able to quantitatively verify the intuition that stream processing platforms will offer better SLAs that the current batch design, and offer a meaningful, real-world workload to verify fast-data platforms. But the caveats mentioned above in operationalizing such an architecture limit their immediate impact within \uid.





%
\section{Conclusions}
\label{sec:conclusions}
In this article, we have proposed a Big Data benchmark for high velocity, \emph{fast-data} applications based on an \emph{eGovernance} workload. This addresses a gap in existing benchmarks that are based on web or enterprise applications, and are often volume-oriented. 

Some characteristics of the workload stand out. The enrollment dataflow's input messages have a larger size, atypical of event streams for fast-data platforms that tend to be in the order of KBs in size. The bi-modal distribution of data streams for both the workloads is seen in event streams from other domains too, and is consistent with human activity patterns that are intrinsic to eGovernance platforms. The enrollment dataflow also has control-flow built-in, when packets fail validation and take a different path, and this  impacts the probability with which different paths are taken. This is captured by the outgoing edge's selectivity. Both of these variations impact resource utilization, and motivate the need for elastic resource provisioning. Better resource allocation models for determining the degree of parallelism per task will also be valuable, and help meet the SLAs while conserving resources.

The \uid workload is unique in its scale, being the largest of its kind in the world, but the characteristics of this workload can be seen in other public sector services such as the Department of Motor Vehicles or the Passport Office issuing IDs, and hence generalizable. Offering a benchmark based on these eGovernance workloads allows us to validate Big Data platforms for these socially important services, and offers the research and practitioner community additional transparency into the internal working of such mission-critical operations. While these results can also help improve UIDAI's Big Data architecture and their SLAs, the practical limitations discussed earlier stand.

Besides the many extensions to this work that was discussed earlier, it is worth examining other such public sector workloads to understand features are intrinsic to them, and set them apart from enterprise workloads. This will help design more effective Big Data solutions and potentially open up research opportunities.






%
\section*{Acknowledgments}
We acknowledge inputs provided by Dr. Vivek Raghavan from UIDAI, and UIDAI's public reports in preparing this article. The views and opinions of authors expressed herein do not necessarily state or reflect those of the Government of India or any agency thereof, the UIDAI, nor any of their employees.


%
%
\bibliographystyle{splncs03}

\bibliography{paper}

\clearpage
\end{document}